# INFLUENCE OF MICRO-CANTILEVER GEOMETRY AND GAP ON PULL-IN VOLTAGE

*W.F. Faris\*, H. M. Mohammed\*\*, and M. M. Abdalla\*\*\*, C. H. Ling\*\**

\*Department of Mechanical Engineering, Faculty of Engineering
International Islamic University Malaysia, P.O. Box 10, 50728 Kuala Lumpur, Malaysia
waleed@iiu.edu.my

\*\*School of mechanical Engineering , University of Nottingham Jalan Broga43500 Semenyih,
Selangor Darul Ehsan Malaysia
Hazem.Dermdash@nottingham.edu.my

\*\*\*Department of Aerospace Engineering, College of Engineering
Technical University of Delft, Delft, Netherland
M.M.Abdalla@lr.tudelft.nl

**ABSTRACT**

In this paper, we study the behaviour of a micro-cantilever beam under electrostatic actuation using finite difference method. This problem has a lot of applications in MEMS based devices like accelerometers, switches and others.

In this paper, we formulated the problem of a cantilever beam with proof mass at its end and carried out the finite difference solution. we studied the effects of length, width, and the gap size on the pull-in voltage using data that are available in the literature. Also, the stability limit is compared with the single degree of freedom commonly used in the earlier literature as an approximation to calculate the pull-in voltage.

## 1. INTRODUCTION

Microsensors represent a large section of microsystems' market. Microsensors offer the advantage of replacing conventional sensors in a one to one fashion while saving weight, energy and cost [3].

Several studies have investigated the behaviour of electrostatically actuated microbeams in microsensors. Yang et al.[9] modelled a clamped-clamped beam with length 350 μm. The model is test by passing four steps electrostatic voltage of 21V, 22V, 25V and 30V. The model is solved by finite differential method (FDM) then compared with the results from the reduced model generated by Karhunen-Loeve/ Galerkin approach. The macromodel method saves more time (about 502.1 speed up factor) and only contribute about 0.9% error.

Hu et al.[5] solved a model of microcantilever beam with analytical Reileigh Riz method. The purpose of this paper is to verify the validity of neglecting the higher terms in the electrostatic force term. The result shows that it only valid when the applied voltages are below the pull in voltage. Hu also do a dynamic analysis of the beam subjected to an AC bias using Runge Kutta method to solve ordinary differential equation. Hu found that the resonant frequencies decrease with the increasing magnitude of applying voltage. Nayfeh and Younis[8] presented a new approach to the modelling and simulation of flexible microstructures under the effect of squeeze-film damping. They applied perturbation methods to the compressible Reynolds equation and solved the equation using finite element method. The results compared to which get from experiment are acceptable. The further analysis was done on fully clamped and clamped-free-clamped-free plate. Abdel-Rahman et al.[1] modelled a plate clamped at both ends and free along its width. They used shooting method to solve and compared the results from previous results. There are good agreement between them. At last, they concluded that one may use $\alpha_1$ (6 x gap distance/thickness) or axial force to stiffen the microbeam. However, only $\alpha_1$ are used to tune the relationship between the natural frequency and voltage. Collenz et al.[2] developed an alternative approach based on a sequential field-coupling (SFC) algorithm to deal with strongly non-conservative electrostatic loads. Collenz modelled a cantilever beam with length to gap ratio of one (l/g =1). Collenz found that pull in cannot occur because the gap is as big as the beam length, therefore the beam tip cannot touch the underlying plane. Hung and Senturia [6] focused on obtaining macromodels based on global basis functions generation from an approach that is mathematically equivalent to Karhunen-Loeve/ Galerkin analysis of a small but representative ensemble of dynamic FEM runs. They found that macromodel speed up simulation by a factor of 37 over a FDM with less than 2% error.

Faris and Abdalla [4] used Galerkin approximation approach to solve MEMS based sensor under thermal loading.

Finite difference was not exploited in the solution of MEMS based problems, though in our opinion it is quite effective for a domain of problems which are easily formulated mathematically.

## 2. MODELING OF A MICRO-BEAM

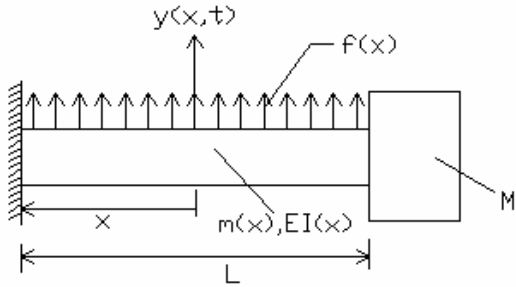

Figure 1 Cantilever Beam with a Lumped Mass at the End

Our model is a cantilever beam with a proof mass suspended at the end as shown in Figure 1. We apply Hamilton principle in developing models to analyze beam behaviour. Hamilton principle is a consideration of the motion of an entire system between two times, $t_1$ and $t_2$. The extended Hamilton's principle:

$$\int_{t_1}^{t_2}(\delta T - \delta V + \delta W_{nc})dt = 0, \quad \delta y(x,t) = 0, \quad 0 \le x \le L, \quad t = t_1, t_2 \qquad (1)$$

We start derive the beam bending equation by write the kinetic energy expansion of the beam:

$$T = \frac{1}{2}\int_0^L m(x)\left[\frac{\partial y(x,t)}{\partial t}\right]^2 dx + \frac{1}{2}M\left[\frac{\partial y(L,t)}{\partial t}\right]^2, \qquad (2)$$

and the strain energy (internal potential) is stated as:

$$V(t) = \frac{1}{2}\int_0^L EI(x)\left[\frac{\partial^2 y(x,t)}{\partial x^2}\right]^2 dx, \qquad (3)$$

The virtual work of the nonconservative distributed forces is expressed as:

$$\overline{\delta W_{nc}}(t) = \int_0^L f(x,t)\delta y(x,t)dx \qquad (4)$$

Substituting into Hamilton's principle (1), we obtain:

$$-\frac{\partial^2}{\partial x^2}\left[EI(x)\frac{\partial^2 y(x,t)}{\partial x^2}\right] + f(x,t) = m(x)\frac{\partial^2 y(x,t)}{\partial t^2}$$
(5a)

and the boundary conditions at $x = 0$

$$y(x,t) = 0, \qquad \frac{\partial y(x,t)}{\partial x} = 0, \qquad (5b)$$

and at $x = L$

$$EI(x)\frac{\partial^2 y(x,t)}{\partial x^2} = 0, \qquad \frac{\partial}{\partial x}\left[EI(x)\frac{\partial^2 y(x,t)}{\partial x^2}\right] - M\frac{\partial^2 y(x,t)}{\partial t^2} = 0, \qquad (5c)$$

In the rest of the paper, we assume the $I(x)$ and $m(x)$ are constant over the length of the beam.

It is our interest to consider the effect of electrostatic forces on the response of the microbeam. The electrostatic load depends on the beam deflection as:

$$f(x) = \frac{\varepsilon b V^2}{2(G-y)^2} \qquad (6a)$$

Where: $E$ = Young Modulus, $m$ = mass per unit length of beam, $I = bh^3/12$ = cross section area moment of inertia, $b$ = width of beam, $h$ = height of beam, $G$ = gap distance, $\varepsilon$ = free space permittivity or dielectric constant of vacuum, $t$ = time.

As a result, equation (5) can be expressed as:

$$-EI\frac{\partial^4 y(x,t)}{\partial x^4} + \frac{\varepsilon b V^2}{2(G-y)^2} = m\frac{\partial^2 y(x,t)}{\partial t^2} \qquad (6b)$$

## 3. SYSTEM GOVERNING EQUATIONS

We consider a cantilever beam with a proof mass suspended at the end, actuated by an electrostatic force which consists of a DC component, $V_p$ and an AC component $v(t)$. We assume that the transverse deflection of beam, $y$, is constant along the width of beam. The beam equation and its boundary conditions at $x = 0$ and at $x = L$ can be expressed as:

$$EI\frac{\partial^4 y(x,t)}{\partial x^4} + m\frac{\partial^2 y(x,t)}{\partial t^2} = \frac{\varepsilon b(V_p + v(t))^2}{2(G-y)^2} \qquad (7)$$

$$y(x,t)=0, \qquad \frac{\partial y(x,t)}{\partial x}=0, \qquad x=0 \qquad (7a)$$

$$EI\frac{\partial^2 y(x,t)}{\partial x^2}=0, \qquad EI\frac{\partial^3 y(x,t)}{\partial x^3}-M\frac{\partial^2 y(x,t)}{\partial t^2}=0, \qquad x=L \qquad (7b)$$

Where: $E$ = Young Modulus, $m$ = mass per unit length of beam, $I = bh^3/12$ = cross section area moment of inertia, $b$ = width of beam, $h$ = height of beam, $G$ = gap distance, $\varepsilon$ = free space permittivity or dielectric constant of vacuum, $t$ = time.

The microbeam deflection under an electric force is composed of static component due to the DC voltage, termed as $y_s(x)$ and dynamic component due to the AC voltage, termed as $u(x,t)$ that is:

$$y(x,t) = y_S(x,t) + u(x,t) \quad (8)$$

As a result, the beam equation can be modified as

$$EI\left[\frac{\partial^4 y_s}{\partial x^4} + \frac{\partial^4 u}{\partial x^4}\right] + m\frac{\partial^2 u}{\partial t^2} = \frac{\varepsilon b(V_p + v(t))^2}{2(G - y_s - u)^2} \quad (9)$$

To calculate static deflection, $y_s$, we put time derivation and the AC forcing term in equation (9) equal to zero and obtain

$$EI\frac{\partial^4 y_s}{\partial x^4} = \frac{\varepsilon b V_p^2}{2(G - y_s)^2} \quad (10)$$

$$y_s = 0, \quad \frac{\partial y_s}{\partial x} = 0, \quad \text{at } x=0 \quad (10a)$$

$$EI\frac{\partial^2 y_s}{\partial x^2} = 0, \quad EI\frac{\partial^3 y_s}{\partial x^3} - M\frac{\partial^2 y(x,t)}{\partial t^2} = 0, \quad \text{at } x = L \quad (10b)$$

To solve the eigenvalue problem, we set AC forcing term in equation (9) equal to zero and use equation (10) to eliminate the term representing equilibrium position.

$$\frac{\varepsilon b V_p^2}{2(G - y_s)^2} + EI\frac{\partial^4 u}{\partial x^4} + m\frac{\partial^2 u}{\partial t^2} = \frac{\varepsilon b V_p^2}{2(G - y_s - u)^2} \quad (11)$$

Expanding the nonlinear electrostatic force term by Taylor series with respect to equilibrium position, $u = 0$, gives

$$\frac{\varepsilon b V_p^2}{2(G - y_s - u)^2} = \frac{\varepsilon b V_p^2}{2}\left[\frac{1}{(G - y_s)^2} + \frac{2}{(G - y_s)^3}u + \frac{3}{(G - y_s)^4}u^2 + ...\right] \quad (12)$$

Based on the small displacement assumption, the higher order terms can be neglected, thus the nonlinear electrostatic force can be linearized as

$$\frac{\varepsilon b V_p^2}{2(G - y_s - u)^2} = \frac{\varepsilon b V_p^2}{2}\left[\frac{1}{(G - y_s)^2} + \frac{2}{(G - y_s)^3}u\right] \quad (13)$$

Substituting equation (13) into equation (11) results in

$$EI\frac{\partial^4 u}{\partial x^4} + m\frac{\partial^2 u}{\partial t^2} - \frac{\varepsilon b V_p^2}{(G - y_s)^3}u = 0 \quad (14)$$

.

**3. FINITE DIFFERENCE SOLUTION**

Refer to the grid below, by assuming that at $x = 0$, $i = 3$ and at $t = 0$, $j = 4$. Hence, for $j \leq 4$, $y_{i,j} = 0$ since the beam is not deflected at $t \leq 0$.

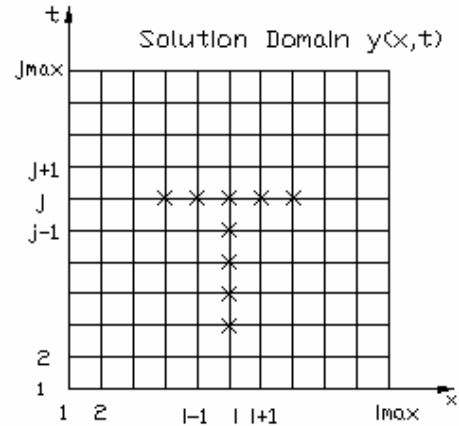

Figure 2 Grid Point for the Dynamic Behavior Proble

$$-\frac{EI}{h^4}\left[y_{i-2,j} - 4y_{i-1,j} + 6y_{i,j} - 4y_{i+1,j} + y_{i+2,j}\right] + \frac{\varepsilon b v_t^2}{2(G - y_{i,j})^2}$$

$$= \frac{m}{12k^2}\left[11y_{i,j-4} - 56y_{i,j-3} + 114y_{i,j-2} - 104y_{i,j-1} + 35y_{i,j}\right] \quad (15)$$

At $x = 0$, $\quad y_{i,j} = 0;$ (15a)

$$y_{i-2,j} - 8y_{i-1,j} + 8y_{i+1,j} - y_{i+2,j} = 0 \quad (15b)$$

At $x = L$,
$$\frac{EI}{12h^2}\left[-y_{i-2,j} + 16y_{i-1,j} - 30y_{i,j} + 16y_{i+1,j} - y_{i+2,j}\right] = 0 \quad (15c)$$

$$\frac{EI}{2h^3}\left[-y_{i-2,j} + 2y_{i-1,j} - 2y_{i+1,j} + y_{i+2,j}\right]$$
$$= \frac{M}{12k^2}\left[11y_{i,j-4} - 56y_{i,j-3} + 114y_{i,j-2} - 104y_{i,j-1} + 35y_{i,j}\right] \quad (15d)$$

According to the assumption stated above, at $x = 0$, $i = 3$, from the boundary condition (22a) and (22b), we get:

$$y_{3,j} = 0, \quad (16a)$$
$$y_{1,j} - 8y_{2,j} + 8y_{4,j} - y_{5,j} = 0 \quad (16b)$$

and from the beam equation at $x = 0$ or $i = 3$,

$$y_{1,j} - 4y_{2,j} - 4y_{4,j} + y_{5,j} = 0 \quad (16c)$$

Rearrange equation (16), we get the equation for $y_{1,j}$, $y_{2,j}$ and $y_{3,j}$, which are

$$y_{1,j} = 4y_{2,j} + 4y_{4,j} - y_{5,j} \quad (17a)$$

$$y_{2,1} = \frac{1}{8}\left[y_{1,j} + 8y_{4,j} - y_{5,j}\right] \quad (17b)$$

$$y_{3,j} = 0, \quad (17c)$$

Solving equation (15) for $4 \leq i \leq N-2$, we get

$$y_{i,j} = \frac{1}{\left(6\frac{EI}{h^4} + 35\frac{M}{12k^2}\right)} \left\{ \begin{array}{l} \frac{\varepsilon b v_t^2}{2(G-y_{i,j})^2} - \frac{EI}{h^4}\left[y_{i-2,j} - 4y_{i-1,j} - 4y_{i+1,j} + y_{i+2,j}\right] \\ -\frac{m}{12k^2}\left[11y_{i,j-4} - 56y_{i,j-3} + 114y_{i,j-2} - 104y_{i,j-1}\right] \end{array} \right\}$$
(18)

In addition, for $i = N-2$, we have to consider another two boundary condition to find $y_{N-1,j}$ and $y_{N,j}$. From equation (15c) and equation (15d), we get

$$y_{N-1,j} = \frac{1}{16}\left[y_{N-4,j} - 16y_{N-3,j} + 30y_{N-2,j} + y_{N,j}\right] \quad (19a)$$

$$y_{N,j} = \left[y_{N-4,j} - 2y_{N-3,j} + 2y_{N-1,j}\right]$$
$$+ \frac{2Mh^3}{12k^2 EI}\left[\begin{array}{l}11y_{N-2,j-4} - 56y_{N-2,j-3} \\ +114y_{N-2,j-2} - 104y_{N-2,j-1} + 35y_{N-2,j}\end{array}\right]$$
(19b)

To facilitate in programming, we assume that $A = EI/h^4$, $B = \varepsilon b v_t^2/2$, $C = m/12k^2$, $D = M/12k^2$ and $F = EI/2h^3$. So the equations describe $y_{i,j}$ for $j \geq 4$ is expressed below.

$$y_{1,j} = 8y_{2,j} - 8y_{4,j} + y_{5,j} \quad (20a)$$

$$y_{2,j} = \frac{1}{4}\left[y_{1,j} - 4y_{4,j} + y_{5,j}\right] \quad (20b)$$

$$y_{3,j} = 0 \quad (20c)$$

for $4 \leq i \leq N-2$,

$$y_{i,j} = \frac{1}{(6A+35C)} \left\{ \begin{array}{l} \frac{B}{(G-y_{i,j})^2} - A\left[y_{i-2,j} - 4y_{i-1,j} - 4y_{i+1,j} + y_{i+2,j}\right] \\ -C\left[11y_{i,j-4} - 56y_{i,j-3} + 114y_{i,j-2} - 104y_{i,j-1}\right] \end{array} \right\}$$
(20d)

$$y_{N-1,j} = \frac{1}{16}\left[y_{N-4,j} - 16y_{N-3,j} + 30y_{N-2,j} + y_{N,j}\right] \quad (20e)$$

$$y_{N,j} = \left[y_{N-4,j} - 2y_{N-3,j} + 2y_{N-1,j}\right]$$
$$+ \frac{D}{F}\left[11y_{N-2,j-4} - 56y_{N-2,j-3} + 114y_{N-2,j-2} - 104y_{N-2,j-1} + 35y_{N-2,j}\right]$$
(20f)
.

## 4. RESULTS AND ANALYSIS

In this paper, we study the behavior of a microbeam under a nonlinear electrostatic force. First, we present a numerical procedure to solve the static behaviour or boundary value problem of a microbeam under DC electrostatic force. We are interested in finding the pull-in voltages corresponding to static deflection at different beam lengths.

Second, we determine the natural frequencies and mode shapes of microbeam under DC electrostatic force. Equations (10) describe the microbeam static deflection under static electrostatic force. We use finite difference method to solve the problem numerically for $y_s$.

$$EI\frac{\partial^4 y_s}{\partial x^4} = \frac{\varepsilon b V_p^2}{2(G-y_s)^2} \quad (21)$$

$$y_s = 0, \quad \frac{\partial y_s}{\partial x} = 0, \quad \text{at } x=0 \quad (21a)$$

$$EI\frac{\partial^2 y_s}{\partial x^2} = 0, \quad EI\frac{\partial^3 y_s}{\partial x^3} - M\frac{\partial^2 y(x,t)}{\partial t^2} = 0, \quad \text{at } x = L$$
(21b)

We initially set $y_s^0$ as zero and iterate until it converge to a value within 0.1%. Considering a cantilever beam with a mass suspended at the end made of silicon. The geometries and material parameter are given in table 1.

| Parameters | Values |
|---|---|
| Length (L) | 300 µm |
| Width (W) | 50 µm |
| Thickness (t) | 3 µm |
| Initial Gap Distance (G) | 3 µm |
| Young Modulus (E) | 160 x $10^9$ kg/m$^2$ |
| Density (ρ) | 2330 kg/m$^3$ |
| Free Space Permittivity (ε) | 8.8541878 x $10^{-12}$ (F/m) |

Table 1 The Geometric and Material Parameters of the Microbeam

We solve equations (21) for a range of electrostatic forces by changing the applied voltage. Then, we will change the length (L) to see its effect on static with respect to applied voltages ($V_p$).

Figure 3 shows the deflection along the microbeam with dimensions L = 300 µm, W = 50 µm, t = 3 µm, G = 3 µm at voltage = 10 V. The beam is fixed at one end. The more distance the point from the fixed end is, the larger will be the deflection.

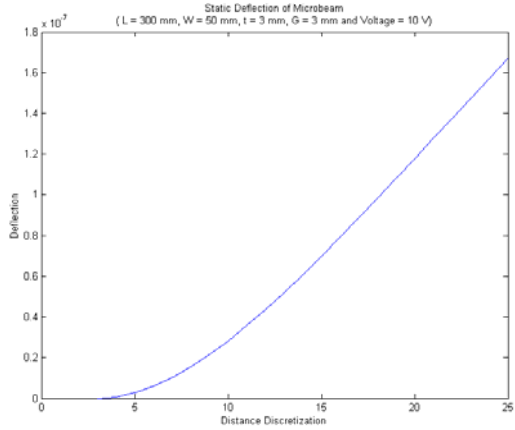

Figure 3   Static Deflection of Microbeam
( L = 300 µm, W = 50 µm, t = 3 µm, G = 3 µm and Voltage = 10 V)

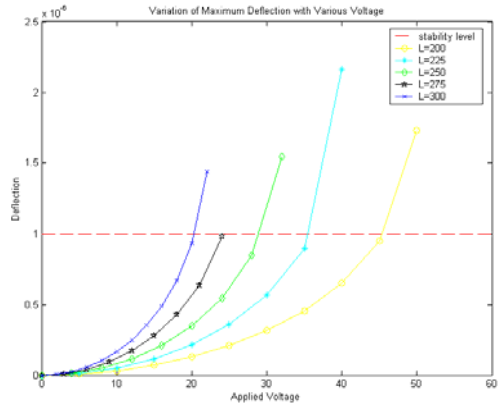

Figure 4 Variation of maximum Deflection with applied Voltage until Pull-In for L=200 µm, 225 µm, 250 µm, 275 µm and 300 µm.

Figure 4 shows the variation of the maximum static deflection of microbeam, $W_{max} = Y_s = y_s(x=L)$, with applied voltages for L=200 µm, 225 µm, 250 µm, 275 µm and 300 µm. Other geometry dimensions are remaining same. The dashed line represents the stability limit, $W_{max} = G/3$ predicted by the single-degree-of-freedom spring-mass model. As applied voltage increases, the maximum deflection increases. The slope of maximum deflection, $\partial y_{max}/\partial V_p$, increases as the voltage increases and finally approach infinity when it reaches pull in voltage. Beside this, maximum deflection is also larger for longer beam. This relationship is linear at low voltage and becomes increasingly nonlinear when the voltage applied is increased. The slope of maximum deflection, $\partial y_{max}/\partial V_p$, is larger for longer beam at specific voltage. As a result, microbeam with shorter length can sustain larger voltage before it collapses.

Variations of the maximum deflection with voltage for various thicknesses are shown in figure 5. As voltage increases, static deflection increases. At specific voltage, maximum deflection increasing as thickness is reduced. The slope of maximum deflection, $\partial y_{max}/\partial V_p$, increases more rapidly for thin beam. This proves that the pull in voltage is larger for thicker beam.

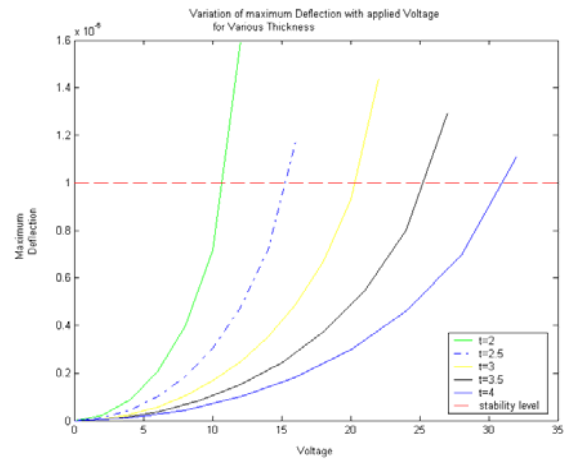

Figure 5 Variation of the Maximum Deflection with Voltage until Pull-In for t=2 µm, 2.5 µm, 3 µm, 3.5 µm and 4 µm.

Gap distance has influence on the static deflection too. As represent in Figure 6, static deflection will be reducing when the gap between microbeam and support is increased.

.

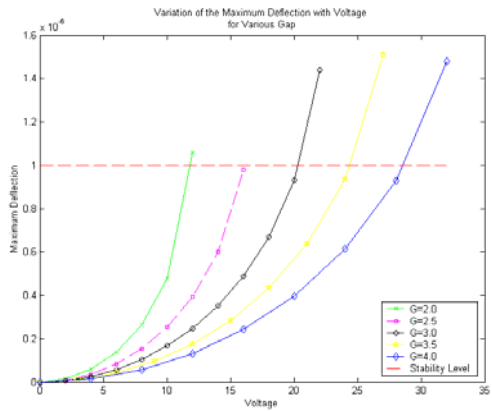

Figure 6    Variation of the Maximum Deflection with Voltage for until Pull-In for G=2 µm, 2.5 µm, 3 µm, 3.5 µm and 4 µm.

## 5. CONCLUSIONS

In this paper, we formulate the problem of a cantilever beam with proof mass at the end which is used in many applications, most notably is MEMS based accelerometer. We used Hamiltonian principle as a tool of this formulation. Also, we explained the solution using finite difference technique. We solved the static part of the problem and decided the pull-in voltage for different geometric parameters of the beam. The parameters studied were beam length, thickness, width, and gap. The finite difference method proved to be effective and easy to code for solving such problems. The length and thickness were found to have a great influence on the pull-in voltage while the width was found of no influence on the pull-in voltage. The gap also was found to have a significant influence on the pull-in results as well.

Also, it was found that the stability limits from the single degree of freedom approximation is quite erroneous compared to full geometry calculations.

## 6. ACKNOWLEDGEMENT

The first author would like to thank the Research Center at IIUM for the financial support to carry out this work..

**Appendix**
**One-Dimensional Pull-In Voltage Analysis**

When a voltage is applied between two electrodes, the electrostatic force between them pulls them together. The restoring force of the beam, arising from the beam's stiffness, resists the electrostatic force. When the voltage is increased, a point is reached where the electrostatic force equals the spring restoring force of the beam. If this point, known as pull-in voltage, is passed, the beam will snap to the substrate.

In one-dimensional geometries, the pull-in phenomenon is greatly simplified and the basic dependencies are easily visible. Now, the displacement field is just a scalar $y$, and the mechanical force, $F_m$ and electric force, $F_e$ are algebraic functions of $u$. The pull-in position is determined by the equality of the forces and their derivatives,

$$F_m = F_e \qquad (1)$$

$$\frac{\partial F_m}{\partial y} = \frac{\partial F_e}{\partial y} \qquad (2)$$

Dividing the two equations we obtain

$$\frac{F_m}{K_m} = \frac{F_e}{K_e} \qquad (3)$$

where we have defined the electric and mechanical spring constants, $K_e = \frac{\partial F_e}{\partial y}$ and $K_m = \frac{\partial F_m}{\partial y}$ respectively and it is valid for all voltage driven electro-mechanical systems. It may equally well be written as

$$y = \gamma \frac{F_e}{F_m} \qquad (4)$$

where $\gamma = y K_m / F_m$ describes the nonlinearity of the elasticity. For linear cases $\gamma = 1$ by definition. Equation (4) provides a starting point for the pull-in iteration scheme.

In the one-dimensional case the electric field has an analytical solution, $E = V/(G-y)$, where $G$ is the initial aperture between the mass and the ground level. The electric energy now yields

$$\mathcal{E} = \frac{1}{2} \varepsilon A \frac{V^2}{G-y} \qquad (5)$$

where $A$ is the area of the capacitor. The electric force and spring constant with fixed potential may be obtained by differentiations,

$$F_e = \frac{\partial \mathcal{E}}{\partial y} = \frac{1}{2} \varepsilon A \frac{V^2}{(G-y)^2}, \text{ and } K_e = \frac{\partial^2 \mathcal{E}}{\partial y^2} = \varepsilon A \frac{V^2}{(G-y)^3}$$
$$(6)$$

Inserting the above formulas in Equation (4) and solving for the pull-in position gives $y = G/3$. The relative displacement of $1/3$ is a characteristic value for this one-dimensional case. Also more complicated cases have values independent of the scale and of the material parameters.

The pull-in voltage can be obtained from equation (1) by equating electric and mechanical spring constants, $K_e = K_m$.

$$\varepsilon A \frac{V_{PI}^2}{(2G/3)^3} = K_m \qquad (7)$$

$$V_{PI} = \sqrt{\frac{8}{27} \frac{G^3 K_m}{\varepsilon A}} \qquad (8)$$